\documentstyle[amssymb,12pt,thmsa,sw20lart]{article}

\input{tcilatex}

\begin{document}

\author{A.R. Tumanyan$^{(*)}$, Yu.L. Martirosyan$^{(*)}$, V.C. Nikhogosyan$^{(*)}$,
N.Z. Akopov$^{(*)}$, \\
Z.G. Guiragossian$^{(**)}$, R.M Martirosov$^{(*)}$, Z.N. Akopov$^{(*)}$\\
\\
\vspace*{-3mm} {\footnotesize $(*)$ Yerevan Physics Institute (YerPhI),
Yerevan, Br. Alikanian St.2, 375036, Republic of Armenia }\\
\vspace*{-3mm} {\footnotesize $(**)$ Guest Scientist at YerPhI}\\
\vspace*{-3mm}}
\title{Asynchronous accelerator with RFQ injection for active longitudinal
compression of accelerated bunches}
\date{}
\maketitle

\begin{abstract}
An asynchronous accelerator is described, in which the principle of its
operation permits the active longitudinal bunch compression of accelerated
proton beams, to overcome the space charge limitation effects of intense
bunches. It is shown that accelerated bunches from an RFQ linac can be
adapted for Asynchronac injection for a multiple of choices in the
acceleration frequencies of the RFQ and the Asynchronac. The offered new
type of accelerator system is especially suitable to accelerate proton beams
for up to $100MeV$ energy and hundreds of $mA$ average current.
\end{abstract}

\section{Introduction}

\smallskip According to its principle of operation, the asynchronous
accelerator (Asynchronac) can be viewed to be a machine in between the
following two cases. The Asynchronac can be viewed as a linear accelerator
wrapped into a spiral, in which the harmonic number of the acceleration
voltage and the equilibrium phase of a bunch relative to the acceleration
field can be changed independently. Also, it can be viewed as a separate
orbit cyclotron (SOC) \cite{Martin}\cite{Trinks}\cite{Brovko} and an
asynchronous cyclotron, which has been described earlier \cite{Tumanyan1}%
\cite{Tumanyan2}.

\smallskip

It is simpler to present the concept of the Asynchronac as a modification of
the parameters and the operating mode of a separate orbit cyclotron, without
changing its structure (see $Figure1$). If $R$, the radius and $f$, the
acceleration voltage frequency in a SOC are selected to be large, such that $%
q$, the harmonic number between acceleration cavities is a large integer at
injection:

\begin{equation}
q=\frac{h}{N_{c}}=\frac{2\pi Rf_{rf}}{N_{c}V}  \label{wave}
\end{equation}

(where, $N_{c}$ is the number of resonators in a cyclotron stage and $V$ is
the speed of particles), $q$ would decrease during acceleration as the speed
of particles increased, presenting the following possibilities.

\smallskip

First, it is possible to restrict the increase of radii from injection to
ejection orbits, by not limiting the increase of the average radius of the
machine. Second, it becomes possible to reduce the strength of the bending
magnetic fields without restrictions. Third, it is possible to increase the
length of drift spaces between bending magnets. And fourth, it becomes
possible to independently set the RF equilibrium phase during the
acceleration process.

\smallskip

Thus, the creation of non-isochronous or asynchronous mode of SOC operation
is provided by having the inter-cavity harmonic number, $q$ discretely
change on integer values as the beam transits through the sectors and turns.
By modifying the magnetic path length in the sectors the hopping of integer $%
q$ values is accomplished. These are the modifications of the parameters and
the operating mode of a separate orbit cyclotron, which convert it to become
an Asynchronac, with additional new useful qualities. One such important
quality is now the ability to axially compress and thus to monochromatize
the accelerated proton or ion beam bunches.

Only at low energies rapid changes occur in the speed of particles. Because
it is now possible to hop over integer values of $q$ in a reasonably sized
accelerator radius, the Asynchronac concept can best be applied at energies
below $100$ $MeV$, to produce average beam currents in the hundreds of $mA$.
Such accelerators, in addition to their stand-alone use for specific
applications, can be more useful as the initial injector stage for the
production of intense bunches, which are injected into higher energy proton
accelerators, such as synchrotrons, linacs or cyclotrons, to produce high
and super-high energy intense beams.

In the present study a valuable implementation of the Asynchronac concept is
based on the multiple use of modern RFQ linacs (generating large currents at
small energy spread) \cite{Schempp}.

\smallskip

\section{Conditions to creat bunch compression in the asynchronac}

\smallskip The creation of bunch compression in an Asynchronac is made
possible by the inherently available independent setting of the RF
acceleration equilibrium phase, from one resonator to another. In this case,
the continuous differential equation for synchrotron oscillations does not
apply. Consequently, the value of the equilibrium phase in each resonator is
determined by the expression:

\begin{equation}
\cos \varphi _{e}=\frac{2\Delta E_{s}\pm \Delta E_{r}}{2U_{n}T_{z}\sin \Psi }
\end{equation}

In equation (2) $\varphi _{e}$ is the acceleration phase for the synchronous
particle in a bunch and $\Delta E_{s}$ is the half width of the natural
energy spread of particles, induced in a bunch, $\Delta E_{r}$ is the full
width of the desired energy spread to be induced in a bunch as it passes
through the periodic resonators.

\smallskip

In an injected bunch, a particle having a higher energy than the energy of
the synchronous particle is found normally at the head of the bunch and
those having lower energy than the central one are normally at the tail. For
this normal case, $\Delta E_{r}$ is used with the plus sign in equation (2).
However, for the reverse situation where particles having higher energy than
the synchronous particle appear at the tail of the bunch, $\Delta E_{r}$ in
equation (2) is used with the minus sign.

$\smallskip $

$U_{n}$ is the amplitude of the acceleration voltage in the n-th resonator.
Using electrical and mechanical methods to regulate each cavity channel, the
details of which will be described in a separate paper, the acceleration
field in any n-th sector resonator's particle orbit channel can be
independently tuned. In equation (2) $2\Psi $ is the full phase width of a
bunch and $T_{z}$ is the transit-time factor determined by

\begin{equation}
T=\frac{\sin \Delta \Phi /2}{\Delta \Phi /2}
\end{equation}

where,

\begin{equation}
\Delta \Phi =\frac{2\pi f_{rf}L_{gap}}{\beta c}
\end{equation}

$f_{fr}$ is the frequency of the acceleration field in resonators, $L_{gap}$
is the full acceleration gap in resonators, $\beta $ is the relativistic
velocity factor, $c$ is the speed of light. Normally $T_{z}$ can be
maintained at a constant value if the acceleration gap in a resonator is
increased in proportion to the speed of the accelerated beam, which in turn
reduces the need to increase the amplitude of the acceleration voltage $%
U_{n} $.

\smallskip

The energy gain of a central particle in the bunch $\Delta E_{e}$ after
exiting a resonator, is given by

\begin{equation}
\Delta E_{e}=U_{n}T_{z}\sin \varphi _{e}
\end{equation}

The energy gain for an edge particle, $a$, located at the head of the bunch
and an edge particle, $b$, located at the tail of the bunch are determined by

\begin{equation}
\Delta E_{a}=U_{n}T_{z}\sin (\varphi -\Psi )
\end{equation}

\begin{equation}
\Delta E_{b}=U_{n}T_{z}\sin (\varphi +\Psi )
\end{equation}

The equilibrium phase for acceleration is set within the limits of

\begin{equation}
0<\varphi _{e}<(\frac{\pi }{2}-\Psi )
\end{equation}

so that the phase of off-center particles always remains below $\pi /2$, and
a Gaussian bunch distribution is maintained.

\smallskip

The bunch duration $\tau _{f}$ at the end of any sector (at the space
located between two adjacent resonators) is given by

\begin{equation}
\tau _{f}=\tau _{s}+\Delta \tau
\end{equation}

\begin{equation}
\Delta \tau =\frac{S_{s}}{c}(\frac{1}{\beta _{b}}-\frac{1}{\beta _{a}})
\end{equation}

where $\tau =\Psi T/\pi $, $S_{s}$ is the orbital path length of particles
in the sector and $T$ is the RF period duration.

\smallskip

In equation (10) if a positive sign is obtained for $\Delta \tau $, which is
when $\beta _{a}>\beta _{b}$, this describes bunch elongation and an
increase in $\tau _{f}$, and if a negative sign is obtained, it describes
axial bunch compression, with a corresponding decrease in the value of $\tau
_{f}$.

\smallskip

The last case is possible only for the reverse particle configuration, which
is when $\beta _{a}<\beta _{b}$. Here the normal positioning of particles in
a bunch is altered. The particles $a$, at the head of a bunch, now have
smaller energy with respect to particles at the center and the tail, a
second condition that supports bunch compression to occur.

\smallskip

If in equation (9) a negative sign is obtained for $\tau _{f}$, it means
that over-compression has occurred, after which the normal distribution of
particles in a bunch will be restored.

\smallskip

If in a bunch, conditions are found by the proper selection of the
equilibrium phase to constantly support the reverse particle distribution,
the bunch duration will constantly decrease and overcome the space charge
effect, which drives the elongation and transverse expansion of the bunch.

\smallskip

From equation (10) it is evident that to obtain large and negative values of 
$\Delta \tau $ it is necessary to have large sector path lengths, $S_{s}$
and a large negative difference in the reciprocals of $\beta $. These are
the third and fourth conditions to achieve bunch compression in the
Asynchronac.

\smallskip

The values of beam injection energy $E_{i}$, initial energy spread $\Delta
E_{s,i}$, injected beam emittance, bunch duration $\tau _{s,i}$, the
injection radius $R_{i}$, the number of acceleration cavities $N_{c}$, the
amplitude of acceleration voltage $U_{n}$ and the RF frequency $f_{rf}$, are
selected during the conceptual design of the accelerator, based on various
technical feasibility considerations. These parameters must $\varphi _{e}$
selected beforehand, based on the simultaneous solution of equations (9),
(10), (5) and (2), to determine the required value of the acceleration
equilibrium phase $\varphi _{e}$, so that it produces bunch compression in a
given sector and a specific orbital turn.

\smallskip

However, the analytical solutions of the combined equations appear
sufficiently cumbersome. Consequently, these cannot be reproduced here, but
are found in the computer codes for the calculation and optimization of
these parameters.

\smallskip

The control algorithm for the steering of bunches from one sector to the
other in the Asynchronac is as follows. As a function of the measured energy
spread $\Delta E_{s,i}$ of the injected beam, the duration of bunches $\tau
_{i}$, the design value of the mean path length of the beam in the first
sector $S_{s1}$ and the selected RF acceleration voltage in the first
resonator, the RF phase is set, such that an equilibrium phase $\varphi _{e}$
is produced on the rising side of the acceleration voltage, which after the
passage of a bunch in the first resonator will cause at once the inversion
of particles to occur.

\smallskip

Also, this particle inversion must be sufficiently intense, to reduce the
duration of bunches $\tau _{f1}$ close to zero at the end of the beam orbit
in the first sector. Thus, there are two cases, one which conserves the
inverted distribution, and the other which induces over-compression and then
restores the normal distribution of particles in a bunch.

\smallskip

For the first case, to obtain a monoenergetic beam in the follow-on second
sector, it is necessary to adjust the path length of particles in the first
sector $S_{s1}$ and to set the equilibrium phase in the second resonator, so
that the acceleration takes place on the falling side of the RF field. For
the second case, the rising side of the RF voltage is used to obtain a beam
of zero energy spread. In this case $\Delta E_{s}$ in equation (2) takes on
a negative sign and $\Delta E_{r}$ is set to zero. In the desired case of
preserving inverse distribution of particles, it is necessary to work only
on the rising side of the RF acceleration field. The choices are determined
by the magnitude of $\Delta E_{s}$.

\smallskip

Hence, at the start of the second sector the beam will be monoenergetic and
have a bunch length equal to $\tau _{f1}$, and at the end of the sector the
bunch duration $\tau _{f2}$ will increase due to the space charge effect. At
the same time this provides the possibility to estimate the effect
experimentally.

\smallskip

The equilibrium phase at the third resonator must be set such that the
acceleration occurs again on the rising side of the RF voltage, to obtain an
inverse particle distribution in bunches, so that at the end of the third
sector the bunch duration $\tau _{f3}$ reduces again to almost zero. Thus,
the process repeats itself, with and without alternating the acceleration
mode on the rising and falling side of the RF voltage. For a more graphic
presentation the numerical modeling in the following section is provided.

\smallskip

The path length of particles in sectors is set by the parameters of the
bending magnetic system, which essentially must change from sector to
sector, to provide the desired values of $q$ and $\varphi _{e}$. To have the
possibility of precise tuning and to relieve the maintenance of different
mechanical tolerances of the accelerator components and their alignment, we
propose to place in the sectors different correction elements. This is in
addition to the magnetic lenses for transverse focusing of the beam and a
number of beam monitors. In particular, wiggler type magnets will be
installed in the straight sections to correct the path length of particles.
Finally, evaluations show that in having large beam orbit separation steps
and in the other features of the Asynchronac, the mechanical tolerances of
accelerator elements and their alignment are relatively relaxed, in the
order of $10^{-3}$.

\smallskip

The important relaxation of tolerances in the Asynchronac is one of its main
advantages, in comparison to other similar accelerator structures, namely,
as compared with the isochronous separate orbit cyclotrons \cite{Martin}\cite
{Trinks}. In these the necessity of strictly maintaining the isochronism of
particle motion reduces to having tight tolerances, which in practice are
difficult to implement. Other important advantages are due to the features
of longitudinal bunch compression and strong transverse focusing. As such,
it is possible to consider the acceleration of bunches at an equilibrium
phase close to $90^{0}$, which in turn, increases the efficiency of
acceleration, decreases the number of turns, decreases the beam losses, and
increases the number of accelerated particles in bunches.

\smallskip

Basically, the Asynchronac's deficiency is the uniqueness or unprecedented
nature of the sector bending magnetic system. This complicates the
standardization of their manufacture and tuning, and somewhat increases the
initial commissioning time and manufacturing cost of the accelerator.
However, some technical innovations already made, essentially facilitate the
solution of these problems. This concern, the fabrication of magnet yokes
from iron sheets with the ability to mechanically change the magnetic
lengths and the remote control of the magnetic alignment in each sector and
turn. The individual feeding of the DC bending magnets and partially the
magnetic focusing lenses is straightforward to implement, using modern
electronics and computers.

\smallskip

The issues of beam transverse focusing in this study are not considered,
since known standard solutions can be utilized, as normally found in strong
focusing synchrotrons. In particular, the separate function periodic
magnetic structure can be of the FODO type. In the Asynchronac the main
difference will be the possibility of having a slowly changing betatron
oscillation frequency, in going from one period to another. This will allow
to compensate the frequency shift of these oscillations, which is due to
different effects, including the space charge effect.

\smallskip

\section{Configuring RFQ beams for injection into cyclotron}

The method of forming short duration bunches from modern high frequency
RFQ's, which produce large current and small energy spread beams, can be
modified to produce longer duration bunches at longer inter-bunch spacing,
that becomes acceptable for injection in the lower frequency Asynchronac.
This technique is based on time compressing the RFQ-produced beam, in which
the compression is completed downsteam, at the point of injection into the
Asynchronac, as described in $Figure2(a)$. $Figure2(b)$ shows the resulting
single longer bunch produced from a train of shorter RFQ-produced bunches.

\smallskip

Our proposed scheme to produce the required beam compression is as follows.
RFQ-produced bunches are initially steering in a RF deflector with a
saw-tooth time varying voltage. The saw-tooth period is equal to the period
of the Asynchronac's driving RF frequency. The sequentially more and more
deflected bunches pass through a $180^{0}$ shaping magnet with different
path lengths, as seen in $Figure2(a)$. After which all RFQ bunches within
the saw-tooth period coincide at the time focal point, producing full
compression of the bunch train into a single longer bunch, at the injection
point of the Asynchronac. The RF deflector and the injection point of the
Asynchronac are located at conjugate points about the $180^{0}$ shaping
magnet.

If the duration of the short bunches in a RFQ is designated by $\tau _{RFQ}$
and the period between RFQ bunches is $T_{RFQ}$, the time-compression of the
bunch train produces a single longer bunch $\tau _{cyc}$ for injection into
the Asynchronac, given by

\begin{equation}
\tau _{cyc}=m\tau _{RFG}
\end{equation}

\smallskip where the period between injected bunches will be

\begin{equation}
T_{cyc}=mT_{RFG}
\end{equation}

in which m is the number of RFQ bunches in a train length equal to the
period of the driving saw-tooth ramped voltage.

\smallskip

The path length of any k-th bunch in the train, starting from the RF
deflector up the time-focused injection point of all the idealized paths, is
obtained by

\begin{equation}
L_{k}=2[\frac{\text{a}}{\cos \alpha _{k}}+R(1-\cos \alpha _{k}+\frac{\pi }{2}%
-\alpha _{k})-\text{a}tg\alpha _{k}+b]
\end{equation}

under the conditions of

\begin{equation}
R\geq (\text{a}tg\alpha _{\max })\text{ }and\text{ }b\leq (R\cos \alpha
_{\max })
\end{equation}

The significance of the quantities $R$, a, $b$, $\alpha $ are exhibited in
the geometry of $Figure2(a)$.

\smallskip

The maximum path length of particles will be

\begin{equation}
L_{\max }=2[\text{a}+b+(\frac{\pi R}{2})]
\end{equation}

while the minimum path length is

\begin{equation}
L_{\min }=2[\frac{\text{a}}{\cos \alpha _{\max }}+R(\frac{\pi }{2}-\alpha
_{\max })]
\end{equation}

The separation of maximum and minimum path lengths, under the optimum
condition of

\begin{equation}
R=\text{a}tg\alpha _{\max }\text{ }and\text{ }b=R\cos \alpha _{\max }
\end{equation}

will be

\begin{equation}
\Delta L_{\max }=2R[ctg\alpha _{\max }+\cos \alpha _{\max }+\alpha _{\max
}-co\sec \alpha _{\max }]
\end{equation}

However, from primary considerations, the separation of maximum and minimum
path lengths is

\begin{equation}
\Delta L_{\max }=\beta cT_{cyc}
\end{equation}

Knowing the value of $\Delta L_{\max }$ from equation (19) and inverting
equation (18) produces the optimal turning radius $R_{tr}$ of beam tracks in
the time compression shaping magnet, whereby the overall dimensions and the
magnetic field strength are obtained. Thus, the optimum bending radius is
given by

\begin{equation}
R_{tr}=\frac{0.5\Delta L_{\max }}{ctg\alpha _{\max }-\cos ec\alpha _{\max
}+\cos \alpha _{\max }+\alpha _{\max }}
\end{equation}

\smallskip

\section{Results of numerical calculations}

A numerical example is worked out to present the key performance features
and to indicate a rough cost estimate of the Asychronac. The following
parameters are used in the calculation of the numerical example accelerator
model. The RFQ linac's RF system operates at $350$ $MHz$, producing a proton
beam of $2.0$ $MeV$ energy, an energy spread of $\Delta E_{s}=2\%$ and a CW
current of up to $100$ $mA$. The frequency of the acceleration voltage in
the Asynchronac is chosen to be $50$ $MHz$, i.e. to have a seven-fold
difference in the frequencies of the acceleration fields, between the RFQ
and the Asynchronac. This means that the number of RFQ bunches to be
compressed into a single bunch is $m=7$.

However, in our example calculation, in order to be able to use two funneled
RFQ's for injection, we have assumed 14 bunches to be compressed into a
single bunch for injection, and for the maximum steering angle of the RF
deflector, $\alpha _{\max }=20^{0}$ is selected.

Whereby, the following parameter values are obtained

\smallskip

$\tau _{cyc}=4.0ns$ $T_{cyc}=40.0ns$ $\Delta L_{max}=78.2cm$

$R\thickapprox 35.15cm$ $b\thickapprox 33.0cm$ a$=96.6cm$

\smallskip

The total maximal path length including the magnetic shaping structure will
be $330$ $cm$, and the track length up to the middle of the first resonator
will be approximately $\ 4.4$ $m$. The duration of bunches at the end of the
total path length will increase due to the beam's energy spread $\Delta
E_{s} $, by approximately $2.2$ $ns$, so that the bunch length in the first
resonator of the Asynchronac will be

$\tau _{cyc}=4.0+2.2=6.2ns$

\smallskip

In the given example of the Asynchronac, operating at a RF acceleration
frequency of $50$ $MHz$, the inter-bunch separation is $T_{cyc}=20$ $ns$. To
match with this spacing, the use of two RFQ's will be required, each
injecting at an inter-bunch spacing of $40$ $ns$, which when initially
combined in a RFQ funnel \cite{Johnson}, will yield the required $20$ $ns$
inter-bunch spacing.

\smallskip

Incidentally, the Asynchronac geometry permits further increasing the number
of injector RFQ linacs in a manner analogous to the conventional method of
multi-turn injection. RFQ's with own bunch-train compressors can inject
beams at each sector of the first or subsequent turn. However, each must
have a different injected beam energy that matches the orbit's energy at the
point of injection. Thus, successive RFQ's must have correspondingly higher
beam energies.

\smallskip

The bunch-train-compressed beam from a RFQ injector, through the $180^{0}$
bending magnet enters the Asynchronac's first resonator. $Figure$ $1$
schematically depicts the Asynchronac structure and the beam orbits, only
for the central particles of the first three turns. The orbit radius at
injection is $R_{i}=3.0m$, the orbit-to-orbit separation is $\Delta R=25cm$
and the number of resonators is $N_{c}=4$. The number of sectors is also
equal to $N_{s}=4$ per beam turn, and since the number of turns is $17$, the
number of independent channels and magnets is $4\cdot 17=68$.

\smallskip

The key design parameters of the Asynchronac for the numerical example are
summarized in $Table$ $1$. Room temperature resonators are used in the
design, which increase the machine's radius. Following the development and
operation of modern cyclotron resonators at the Paul Scherrer Institute \cite
{Proc.} and the related designs and models \cite{Fietier}, the operation of
these resonators at RF frequencies of $40-50$ $MHz$ and peak voltages of up
to $1.1$ $MV$ can be made available. These resonators have a length of
approximately $6$ $m$, height of $3$ $m$ and width along the beam of $0.3$ $%
m $, and provide a radial operating clearance for orbits of up to $4$ $m$.

\begin{center}
\smallskip \smallskip \smallskip \smallskip

Table 1. Key Parameter Values of an Asynchronac

\smallskip

\begin{tabular}{llll}
& PARAMETER & UNIT & VALUE \\ 
& Beam Spacie &  & Proton \\ 
E$_{i}$ & Injected Beam Energy & $MeV$ & $2.0$ \\ 
E$_{e}$ & Extracted Beam Energy & $MeV$ & $50.0$ \\ 
R$_{i}$ & Injected Beam Radius & $m$ & $3.0$ \\ 
R$_{e}$ & Extracted Beam Radius & $m$ & $7.0$ \\ 
N$_{c}$ & Number of Acceleration Cavities &  & $4$ \\ 
N$_{m}$ & Number of Sector Magnets &  & $66$ \\ 
H & Field Strength in Sector Magnets & $T$ & $0.11-0.85$ \\ 
$\Delta $E & Energy Gain per Turn & $MeV$ & $0.04-3.60$ \\ 
$\Delta $R & Orbit Turn-to-Turn Separation & $cm$ & $25.0$ \\ 
n & Number of Turns &  & $16.5$ \\ 
h & Harmonic Number &  & $52-58$ \\ 
L$_{m}$ & Length of Sector Magnets & $m$ & $0.75-5.13$ \\ 
L$_{f}$ & Length of Drift Spaces & $m$ & $2.2-9.7$ \\ 
$\tau _{f}$ & Duration of Bunches & $ns$ & $6.2-0.5$ \\ 
N$_{0}$ & Number of Protons per Bunch &  & $2.5\cdot 10^{10}$%
\end{tabular}
\end{center}

\smallskip

Thus, an injected bunch duration of $\ 6.2$ $ns$ is compressed down to$2.5$ $%
ns$ at the end of the first sector, using the parameter values of $U_{n}=130$
$KeV$, $T_{z}=0.95$, $\varphi =55.8^{0}$ and setting the equilibrium phase
in the first resonator at $\varphi _{e}=18.4^{0}$. Under these conditions,
particles at the head of the bunch will have energy equal to $2.104$ $KeV$,
while at the tail of the bunch, particle energy will be $1.975$ $KeV$.
Particles at the equilibrium phase will have an energy of $2.039$ $KeV$ and
the energy spread of the bunch will be $\Delta E_{s}=2.104$ $KeV$ Next,
particle inversion will take place. At the end of the first sector, $\Delta
\tau $ will be $8.9$ $ns$, consequently $\tau _{f}=6.2-8.9=-2.7$ $ns$,
whereby the inversion has been completed and over-compression has occurred.
The subsequent processes are easier to observe in $Figures$ $3-10$, right up
to the achievement of the final proton beam energy of approximately $50$ $%
MeV $, which is produced after $16.5$ turns in the Asynchronac.

It is observed from the numerical results in these figures, that the process
of effective longitudinal beam compression has terminated after the first
three turns. The duration of bunches has attained an almost stationary value
of about $0.5$ $ns$ and the final energy spread of the beam $\Delta E_{s}$
ends up to be zero in the Asynchronac.

We now roughly estimate the maximum particle population in a proton bunch,
as a function of bunch shortening. In our simplified approach we ignore the
intra-beam scattering and wake field effects on bunch lengthening and the
axial focusing from the RF acceleration cavities. In the proton bunch's rest
frame, the energy spread due to the bunch electric self-field is given by

\begin{equation}
\frac{\Delta P^{2}}{2m}=-e\int E_{z}dz
\end{equation}

where $z$ is the longitudinal coordinate and $\Delta P$ is the momentum
spread. The longitudinal space-charge electric field of the bunch is
obtained as \cite{Wiedeman}

\begin{equation}
E_{z}=-\frac{e}{4\pi \varepsilon _{0}}\frac{1}{\gamma ^{2}}(1+2\ln \frac{b}{a%
})\frac{\partial \lambda (z)}{\partial z}
\end{equation}

where is the absolute dielectric constant, a and b are the radii of the
proton beam and the vacuum chamber, respectively, and $\lambda (z)$ is the
particle linear density in the bunch. Taking a Gaussian distribution for the
bunch linear density

\begin{equation}
\lambda (z)=\frac{N_{0}}{\sqrt{2\pi }\sigma }e^{\frac{z^{2}}{2\sigma ^{2}}}
\end{equation}

with $\sigma =0.7z$ as the standard value of the bunch length, and inserting
expression (22) into equation (21), and after integration within the bounds
of the bunch's initial z$_{i}$ and final $z_{f}$ half-lengths, the following
formula is obtained for the maximum particle population in a bunch

\begin{equation}
N_{0}=\frac{\Delta P^{2}}{2m}\frac{4\pi \varepsilon _{0}}{e^{2}}\frac{\gamma
^{2}}{e^{-\frac{z_{f}^{2}}{2\sigma ^{2}}}-e^{-\frac{z_{i}^{2}}{2\sigma ^{2}}}%
}\frac{\sqrt{2\pi }\sigma }{1+2\ln \frac{b}{a}}
\end{equation}

Using the parameters in the above, and taking for the beam radius $a=0.5$ $%
cm $ and the vacuum chamber radius $b=5.0$ $cm$, the estimated maximum
number of protons per bunch is $N_{0}\thickapprox 5\cdot 10^{11}$.

\smallskip

From these figures it is seen that the magnetic field in each sector and
turn has sufficiently different and not necessarily optimized parameters. To
simplify the numerical calculations we assumed that the bending magnets in
each sector and turn consist of single whole units, instead of being a
number of shorter modular magnets that would serve the required purpose. In
making the conceptual design of a specific Asynchronac, the sector and turn
magnets will be modularized with optimized parameters, such that the
differences among modular magnets will be few, to permit standardizing the
manufacturing process.

\smallskip

Rather low values of $0.1-0.85$ $Tesla$ are required for the magnetic field
strength in each sector and turn. The fabrication of small-sized modular
bending magnets at these field strengths is a standard matter. Should the
beam's vacuum chamber need an aperture full width of as much as $10$ $cm$,
this can be easily accommodated, since the turn-to-turn orbit separations
will all be equal at $\Delta R=25$ $cm$. The lengths of the remaining free
drift spaces in sectors and turns, after the allocation of resonators and
bending magnets, will be more than $2.0$ $m$. This will allow not only to
freely install focusing magnetic elements and beam monitoring apparatus, but
also to provide easily $100\%$ extraction of the beam.

\smallskip

A rough estimation of the cost to build and operate the Asynchronac shows
that the cost per megawatt of proton beam produced from the Asynchronac is
much less expensive by an order of magnitude, as compared to a megawatt of
proton beam produced from the high and super-high energy accelerators.

\smallskip

\smallskip

\section{Conclusion}

The primary objective of this paper is to show that the innovated
accelerator type, which we refer to as the Asynchronac, has sufficient
feasibility features for its implementation, in which an accelerated proton
beam of up to $100$ $MeV$ energy and hundreds of $mA$ current can be
effectively bunch-compressed. Consequently, it is important to expedite the
extension of further multifaceted studies on this concept, to improve the
quality of future machines for scientific and applied applications,
potentially using this alternative.

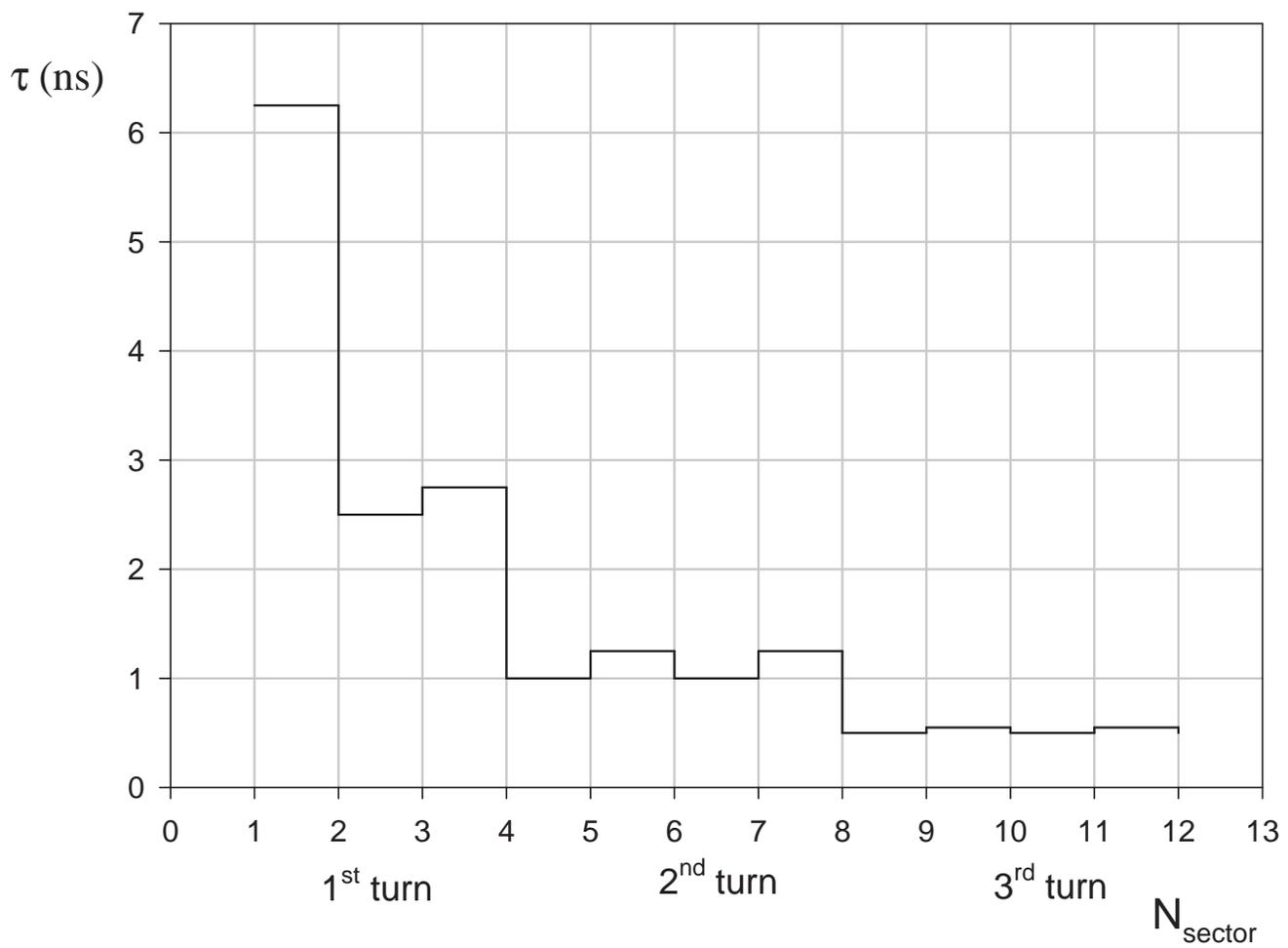
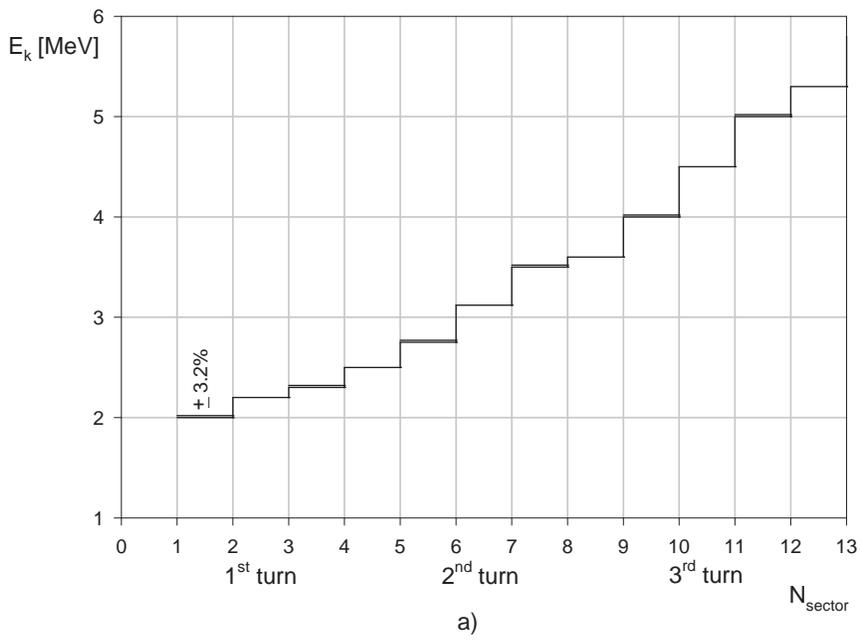
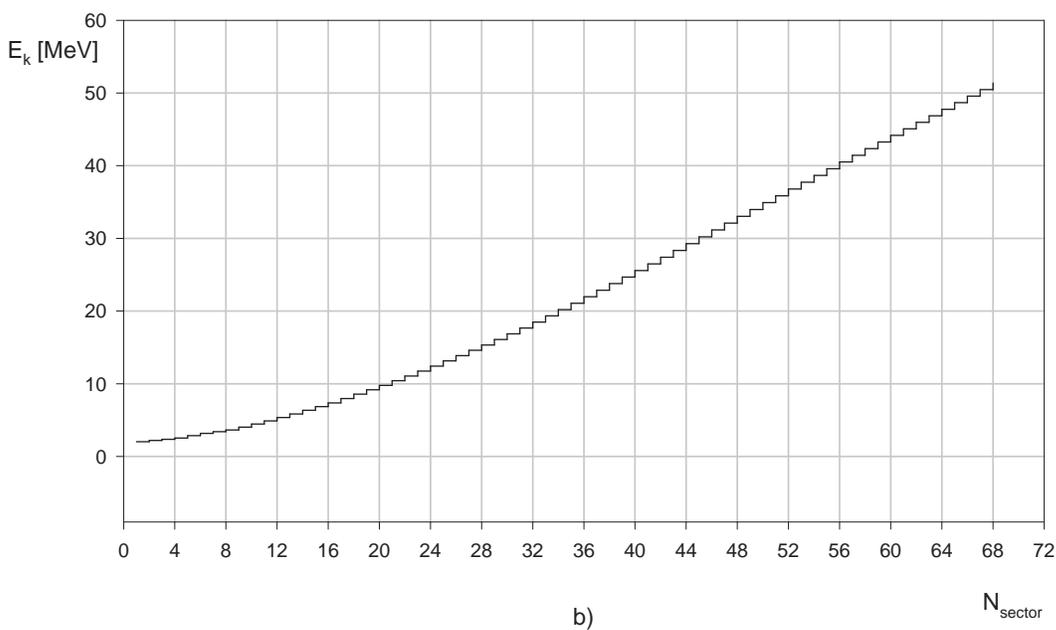
\end{document}